\documentclass[conference,a4paper]{IEEEtran}
\IEEEoverridecommandlockouts
\usepackage{cite}
\usepackage{amsmath,amssymb,amsfonts}

\usepackage{textcomp}
\usepackage{xcolor}
\def\BibTeX{{\rm B\kern-.05em{\sc i\kern-.025em b}\kern-.08em
    T\kern-.1667em\lower.7ex\hbox{E}\kern-.125emX}}

\usepackage{ textcomp }

\usepackage{multirow}
\usepackage{longtable}
\usepackage{amssymb}
\usepackage{graphicx}
\usepackage{xcolor}
\usepackage{algorithm,algorithmic}
\usepackage[algo2e]{algorithm2e}

\begin{document}

\title{A Message Passing Based Receiver for \\ Extra-Large Scale MIMO}
\author{
\IEEEauthorblockN{Abolfazl Amiri, Carles Navarro Manch\'on, Elisabeth de Carvalho  }
\IEEEauthorblockA{ Department of Electronic Systems, Aalborg University, Denmark\\
Email: \{aba,cnm,edc\}@es.aau.dk}
}

\maketitle

\begin{abstract}
We consider a massive MIMO system where the array at the access point reaches a dimension that is much larger than the array  in current systems. 
Transitioning to an extremely large dimension and hence large number of antennas implies  a need to scale up the multi-antenna processing while maintaining a reasonable computational complexity. 
In this paper, we study the receiver of such an extra-large scale MIMO (XL-MIMO) system. 
We propose to base the reception on Variational Message Passing (VMP). The motivation is that the complexity of VMP scales (almost) linearly with the number of antennas and number of users, hence enabling low-complexity reception in crowd scenarios. Furthermore, VMP adapts to the non-stationarities of the MIMO channel that appear due to the large dimension of the array.
Through numerical results, we show significant performance improvement and computational complexity reduction compared to a zero-forcing receiver.
\end{abstract}


\section{Introduction}
Motivated by the significant spectral efficiency gains brought by massive multiple-input multiple-output (MIMO) systems, the study of systems with even larger number of antenna elements is currently under investigation. 
In conventional massive MIMO~\cite{marzetta}, the arrays have a moderate size with an antenna spacing with half the wavelength. In 
extra-large scale MIMO (XL-MIMO), the focus is placed on increasing the dimension of the antenna array at the access point to capture additional spatial degrees of freedom~\cite{xlmimo_mag}. 

\begin{figure}
	\centering
	\includegraphics[scale=0.4,trim={0.6cm 1.6cm 0.6cm 0.6cm },clip]{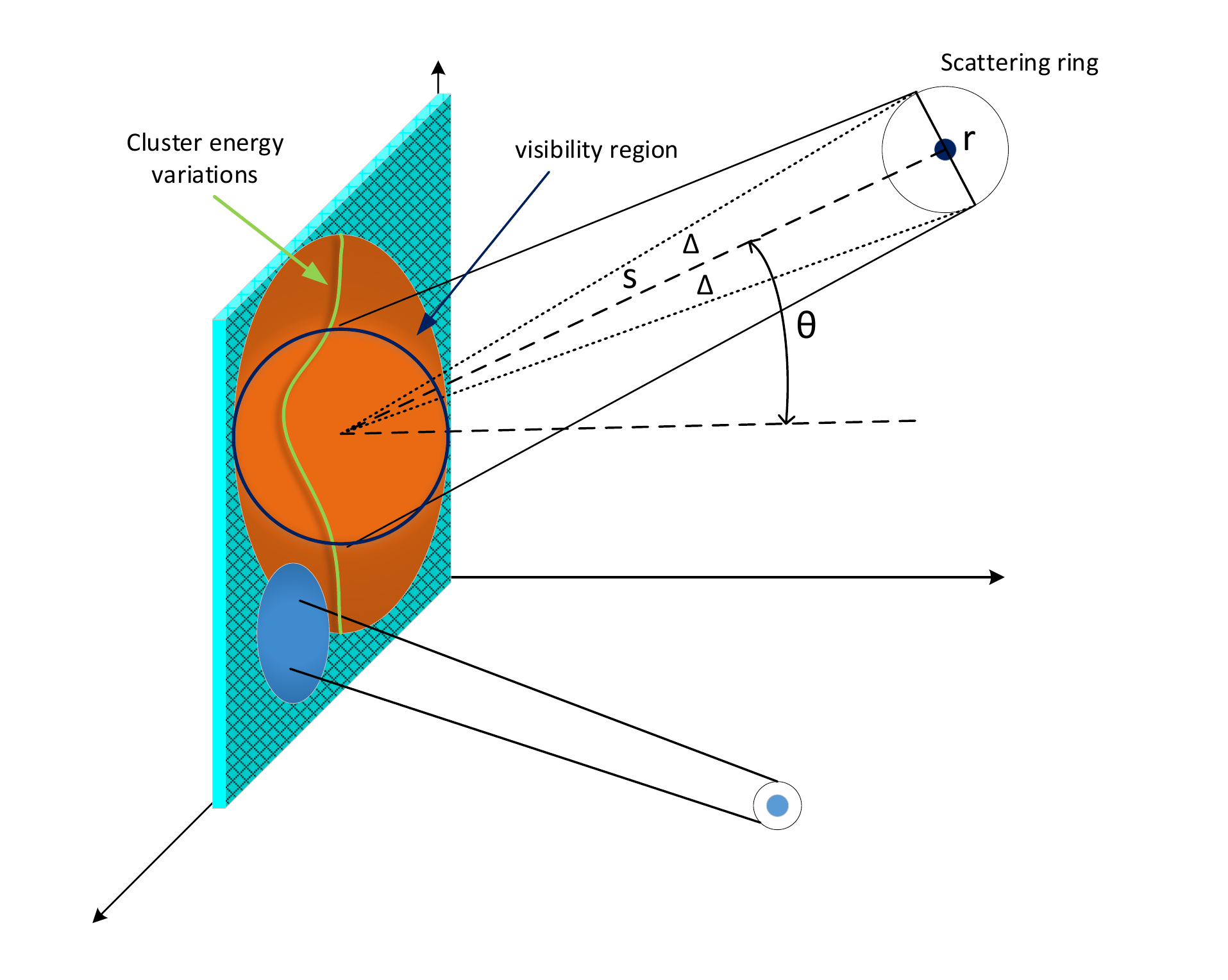}
	\caption{ \small An illustrative example of visibility region and received power distribution over XL-MIMO array.
	}
	\label{fig:ex1}
\end{figure}

Due to the large array dimension and according to the electromagnetic propagation effects, spatial non-wide sense stationary properties appear along the array in XL-MIMO (see Fig.~\ref{fig:ex1}) \cite{xlmimo_mag}. 
This results in the appearance of \textit{visibility regions (VR)}, which are the areas of the array where most of a given users' energy is concentrated.
{The non-stationary properties  impact the performance of XL-MIMO and calls for adapted transceiver designs.}
Another challenge is that of the computational complexity of the receiver processing, especially when the system is serving a large number of users.
Since the conditions for favorable propagation are not satisfied when the number of users is not much smaller than the number of antennas, simple receivers such as a matched filter do not offer good performance~\cite{larsson_energy}. In addition, more advanced linear options such as zero-forcing (ZF) and linear minimum mean squared error (MMSE) receivers have prohibitive complexity due to large matrix inversions. Hence, there is a need for finding multi-user detection algorithms whose complexity scales well with the number of antenna elements and users, and whose performance is comparable to that of classical linear MIMO receivers.

We studied the receiver design for large scale MIMO systems in \cite{xlmimo_GC} where distributed units, called \textit{sub-arrays}, detect users' signals by cooperation. We used a successive interference cancellation (SIC) based method between sub-arrays.
{ Message passing (MP) has been applied to 
 massive MIMO systems. In \cite{som2011low} authors develop low complexity MP methods using graphical models. Both channel estimation and data detection problems with one-bit quantization are solved with MP techniques in \cite{zhang2017one}.}
Recently, authors in \cite{wang2019expectation} used expectation propagation (EP) to solve the symbol detection problem in XL-MIMO systems. They have exploited the sub-array structure to model their EP scheme. 
To the best of our knowledge, the case of heavily loaded XL-MIMO system, where the ratio between the numbers of  BS antennas and active users is small (less than 5), is not studied in the literature. The main challenges in these scenarios are the extreme complexity of the conventional methods and huge degradation in the performance of the benchmark linear receivers. 

In this paper we propose a multi-user symbol detection algorithm for XL-MIMO systems based on variational message passing (VMP). In the VMP framework, the a-posteriori probability of the symbols from all users is approximated by a fully-factorized distribution~\cite{dauwels2007variational,Bishop}, yielding an algorithm with complexity that scales linearly with the number of users and antenna elements, as no matrix inversions are involved. The proposed algorithm is initialized with a maximal ratio combiner (MRC), whose complexity also scales linearly with the system dimensions. In addition, since the variational framework operates with approximations of the posterior probability distributions of the unknown variables rather than with point estimates, it provides an inherent way of optimally fusing the information on a user's symbol obtained from the different antenna elements in the array. This property is especially useful in the presence of spatial non-stationarities and VRs, as occurring in XL-MIMO arrays.

Our complexity analysis shows that the number of multiplications for our proposed algorithm grows much slower with the size of the BS array than for the ZF, especially in crowded scenarios.  We showcase the performance of the proposed receiver in an XL-MIMO system with a large user load, showing that it outperforms the ZF receiver with a significantly lower computational overhead.

\section{System Model}\label{sec:system model}

In this section, we present the system model and introduce a channel model  incorporating spatial non-stationarities.
We denote the number of antennas and simultaneously active users by $M$ and $K$, respectively.
We assume narrow-band transmissions; 
$\mathbf{x}\!\!\in\!\!\mathbb{C}^K$ denotes the vector of complex user symbols, $\mathbf{H}\!\!\!\in\!\!\!\mathbb{C}^{M\times K}$ is the complex channel matrix and $\mathbf{n}\!\!\sim \!\!\mathcal{CN}(0,\sigma_n^2\mathbf{I}_M\!)$ is the AWGN ($\mathbf{I}_M\!$ denotes the identity matrix). We model the received baseband signal $\mathbf{y}\!\in\!\mathbb{C}^{M}$ as:
\begin{align}\label{eq:general_model}
\mathbf{y}=\mathbf{H}\mathbf{x}+\mathbf{n}.
\end{align}
Denoting by $\mathbf{h}_k$ the $k$th column of $\mathbf{H}$, corresponding to user $k$, and adopting the channel  model in \cite{larsson_energy}:
\begin{align}\label{eq:ch_model}
   \mathbf{h}_k=\sqrt{\mathbf{w}_k} \odot\bar{\mathbf{h}}_k,
\end{align}
with $\odot$ denoting the element wise (Hadamard) products between two equal-size vectors.
$\mathbf{w}_k$ captures the effect of large scale fading and has entries
\begin{align}
w_{k,m}={\Omega_k} s_{k,m}^{\nu},\qquad m=1,\dots,M
\end{align}
where $s_{k,m}$ is the distance between user $k$ and BS antenna $m$, $\Omega_k$ is an attenuation coefficient and $\nu$ is the pathloss exponent\cite{Tse:2005:FWC:1111206}.


 $\bar{\mathbf{h}}_k \sim \mathcal{CN}(0,\mathbf{R}_k) $ models a non-line of sight fast fading scenario, with $\mathbf{R}_k$ being a symmetric positive semi-definite channel covariance matrix. Karhunen-Loeve expansion representation of the channel vectors $\bar{\mathbf{h}}_k$ are \vspace{-0.2cm}
  \begin{align}\label{eq:KL_representation}
     \bar{\mathbf{h}}_k=\mathbf{U}_k \mathbf{\Lambda}_k^{\frac{1}{2}}\mathbf{\omega}_k ,
 \end{align}
 where $\mathbf{\omega}_k\in\mathbb{C}^{\zeta\times 1}\sim \mathcal{CN}(0,\mathbf{I})$, $\mathbf{\Lambda}_k$ is an $\zeta\times \zeta$ diagonal matrix with dominant eigenvalues and $\mathbf{U}_k\in\mathbb{C}^{M\times \zeta} $ is the tall unitary matrix of the eigenvectors of $\mathbf{R}_k$
corresponding to the $\zeta$ dominant eigenvalues.

\subsection{Visibility Regions and Correlation Model}\label{sec:non-stationary parameters}
In order to model the channel characteristics in the non-stationary conditions, we refer to the measurement-based data from \cite{gao2013massive}. There, more realistic scenarios exploiting the VRs over a relatively large array ($7.35$ meters) were applied. The authors modeled  different properties for the VRs including VR centers and VR lengths which we denote with $c_k$ and $l_k$, respectively for user $k$. The $c_k$'s are modeled with a uniform random variable over the array, i.e. $c_k\sim \mathcal{U}(0,L)$, where $L$ is the physical length of the XL-MIMO array. The length of the VR follows a log-normal distribution, $l_k\sim \mathcal{LN}(\mu_l,\sigma_l)$. 

Exploiting the well-known \textit{one-ring} model \cite{one_ring} to define $\mathbf{R}_k$,  the correlation between the channel coefficients of antennas $p$ and $q$ is given by\vspace{-0.1cm}
\begin{align}\label{eq:R_def}
    [\mathbf{R}_k]_{p,q}=\frac{1}{2\Delta}\int_{-\Delta}^{\Delta}\exp{\bigl(j\mathbf{f}(\alpha+\theta)(\mathbf{u}_p-\mathbf{u}_q)\bigr)}\text{d}\alpha,\vspace{-0.2cm}
\end{align}
 where $\mathbf{f}(\omega)=-\frac{2\pi}{\lambda}\left(\cos(\omega),\sin(\omega)\right)$ is the wave vector with carrier wavelength of $\lambda$ and $\mathbf{u}_p,\mathbf{u}_q \in \mathbb{R}^2$ are the position vectors of the antennas $p,q$ within the VR of user $k$, angle of arrival of $\omega$ and   $\Delta$ is angular spread which is $\Delta\approx \arctan(\frac{r}{s})$, with $r$ standing for the ring of scatterers radius \cite{rahmati2019energy}.
 Angle $\theta$ is the azimuth angle of user $k$ with respect to antenna array (See Fig.~\ref{fig:ex1}). When either of the antenna indices $p,q$ is outside the VR for user $k$ given by the non-stationary parameters, we have $[\mathbf{R}_k]_{p,q}=0$. In this work, we use a uniform linear array (ULA) configuration and assume a uniform distribution for the scattering rings in front of the array.

\section{Proposed Receiver Algorithm}
In this section we describe our proposed symbol detection method which is based on variational message-passing (VMP). Due to space limitations, we cannot include the full details of the VMP method, and refer the readers to \cite{dauwels2007variational,Bishop} instead. 

\subsection{Variational Message Passing}
\begin{figure}
	\centering
	\includegraphics[scale=0.3,trim={0cm 1cm 0cm 0cm },clip]{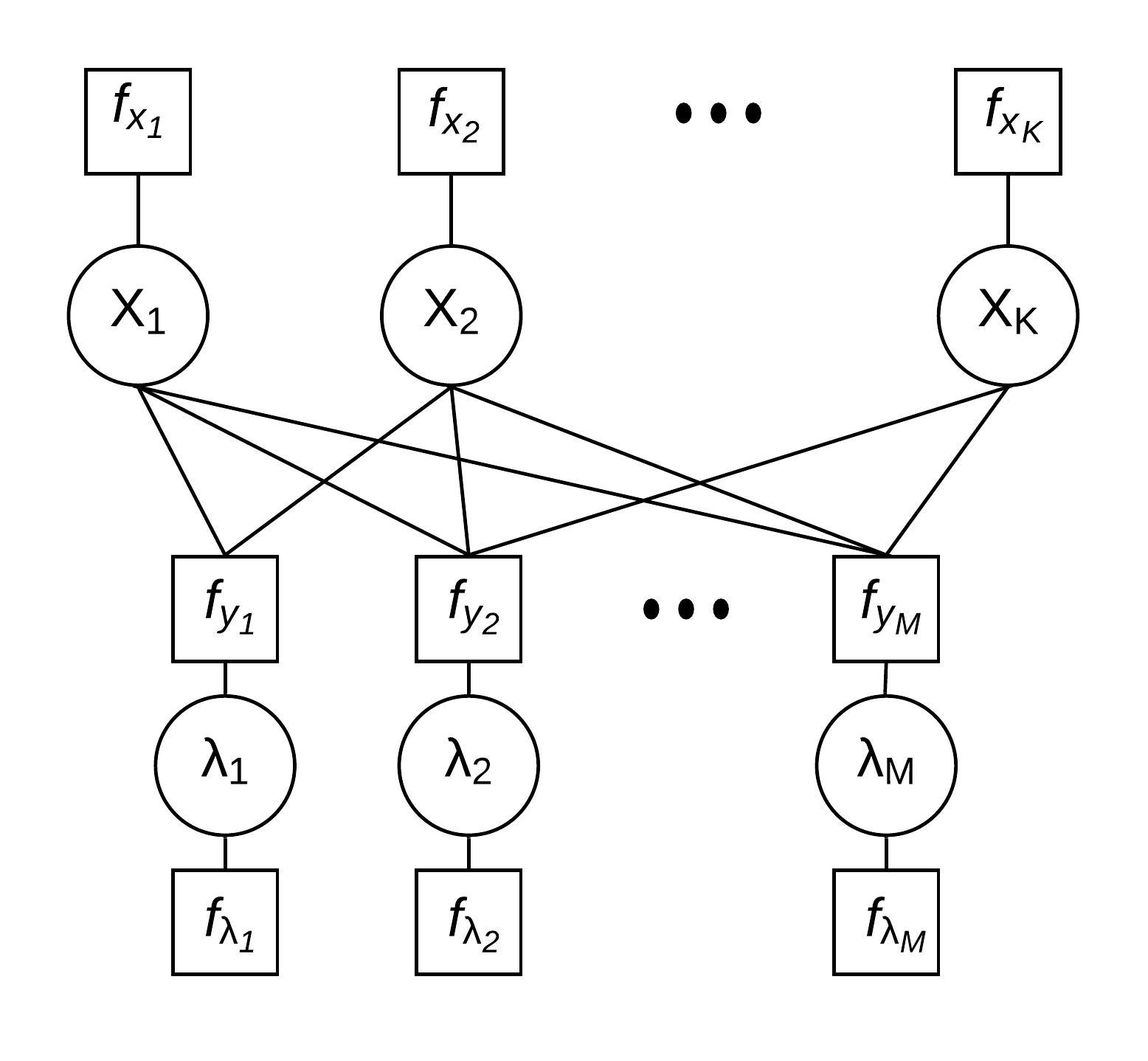}
	\caption{ \small Factor graph representation of the system model.
	}
	\label{fig:ex2}
\end{figure}

We aim to detect the transmitted user symbols ${x}_k$ (for user $k$) which take a value from the constellation set $\mathcal{A}=\{a_1,a_2,\cdots ,a_{|\mathcal{A}|}\}$. Moreover, we also estimate as a nuisance variable the noise precision (i.e. inverse noise variance) $\lambda_b$, $b=1,\dots,M$ at each antenna port. 
The posterior probability density of these two variables factorizes as
\begin{align}
    p&(x_1,\cdots,x_K,\lambda_1,\cdots,\lambda_M|y_1,\cdots y_M)\propto\nonumber\\\label{eq:joint_prob} &\prod_{b=1}^M\underbrace{p(y_b|x_1,\cdots,x_K,\lambda_b)}_{f_{y_b}}\prod_{b=1}^M\underbrace{p(\lambda_b)}_{f_{\lambda_b}}\prod_{k=1}^K\underbrace{p(x_k)}_{f_{x_k}}
\end{align}
The factorization chosen in \eqref{eq:joint_prob} can be visualized in the factor graph representation shown in Fig.~ \ref{fig:ex2}, where variable and factor nodes are illustrated with circles and squares, respectively.

In variational inference/message passing, we approximate the joint posterior of the variables in the system by a fully factorized auxiliary function of the form
\begin{align}\label{eq:auxiliary_function}
    q(\mathbf{x},\lambda_1,\dots,\lambda_M)=\prod_{k=1}^K q_{x_k}(x_k)\prod_{b=1}^M q_{\lambda_b}(\lambda_b)
\end{align}
The individual $q(\cdot)$ factors in the r.h.s. of \eqref{eq:auxiliary_function} are then sequentially updated by minimizing the Kullback-Leibler divergence between the posterior probability function in \eqref{eq:joint_prob} and the approximating auxiliary function with respect to one of the factors at a time. After convergence, they yield approximations of the posterior marginals of the system variables. 

 The variables $\lambda_b$ model the variance of noise $\mathbf{n}$ and residual interference at antenna ports $b = 1, ...., M$. We set their prior as a Gamma distribution \cite{manchon2011receiver}, which is the conjugate prior for the precision of a Gaussian distribution with known mean. Hence, their pdfs read
\begin{align}\label{eq:lambda_gamma}
    f_{\lambda_b}({\lambda_b}) \propto {\lambda_b}^{(\alpha_0-1)}\exp{(-z_0{\lambda_b})}\quad\text{for}\: b\in\{1,\cdots, M\}
\end{align}
where $\alpha_0$ and $z_0$ are respectively the shape and rate parameters. The factors $f_{y_b}$ correspond to the pdf of the signal received at antenna ports $b$ conditioned on the users' symbols and $\lambda_b$, reading
\begin{align}\label{eq:Prob_y}
    p({y}_b|\mathbf{x},{\lambda_b})=\frac{{\lambda_b}}{\pi}\exp (-{\lambda_b}||{y}_b-\mathbf{H}_{[b,:]}\mathbf{x}||^2)
\end{align}
where $\mathbf{H}_{[b,:]}$ denotes $b$th row of $\mathbf{H}$, i.e. the channel for  antenna element $b$. Finally, the prior distribution of the transmitted symbols are modeled as uniform over constellation set $\mathcal{A}$.

To begin with, and according to the definition of the VMP method, the message from factor node $f_{{y}_b}$ to the variable node $\lambda_b$ is
\begin{align}\label{eq:VMP_basic}
   m_{ f_{{y}_b\longrightarrow {\lambda_b}}}({\lambda_b})\propto \exp{(\mathbb{E}_\mathbf{x}\{\ln{\big(p({y}_b|\mathbf{x},{\lambda_b})\big)}\})}
\end{align}
where $\mathbb{E}_\mathbf{x}$ is the expectation 
with respect to the distribution given by $q_{\mathbf{x}}({\mathbf{x}}) = \prod_{k=1}^K q_{x_k}(x_k)$.
After multiple simplifications, we reach to
\begin{align}
     m_{ f_{{y}_b\longrightarrow {\lambda_b}}}({\lambda_b})&\propto {\lambda_b} \exp (-{\lambda_b} Z)
\end{align}
where $Z={||{y}_b-\sum_k \mathbf{h}_k \mu_{{x}_k}||^2+ \sum_k \sigma^2_{x_k}\mathbf{h}_k^H \mathbf{h}_k}$ with $\mu_{x_k}$ and $\sigma^2_{x_k}$ standing for mean and variance of $x_k$, which is derived below.
Now, we can calculate the approximate marginal probability distribution of $\lambda_b$ at antenna port $b$ by multiplying the messages entering the variable node $\lambda_b$ as
\begin{align}\label{eq:q_lambda}
    q_{\lambda_b}\!({\lambda_b})\!\!=\!f_{\lambda_b} \!\times\!  m_{ f_{{y}_b\rightarrow {\lambda_b}}}\! \!=\! {\lambda_b}^{\overbrace{1\!+\!\alpha_0}^{\alpha-1}}e^{ -{\lambda_b}\! \overbrace{(z_0\!+\!Z)}^\beta}
\end{align}
Next, we consider the messages from each antenna to the symbol variables, $m_{f_{y_b}\longrightarrow x_k}$ which is
\begin{align}
    m_{f_{y_b\longrightarrow x_k}}\!\!\! \propto  \mathcal{CN}\bigg(\!x_k;\! \frac{H_{b,k}}{|H_{b,k}|^2}
    ({y}_b\!-\!\!\!\!\sum_{k'\neq k}\!\mu_{x_{k'}}H_{b,k'}),\! \frac{1}{\mu_{\lambda_b}|H_{b,k}|^2}\bigg)\nonumber
\end{align}
where $\mu_{\lambda_b}$ is the mean value of the ${\lambda_b}$ variable which can be calculated from 
\eqref{eq:q_lambda}  as
\begin{align}\label{eq:lambda_bar}
        \mu_{\lambda_b}|_{\substack{\alpha_0=0\\z_0=0}} =\! \frac{\alpha}{\beta}\!
    =\!\frac{1}{{|{y}_b\!-\!\sum_k H_{b,k} \mu_{{x}_k}|^2\!+\! \sum_k \sigma^2_{x_k} |H_{b,k}|^2}}
\end{align}
with $H_{b,k}$ denoting the channel between antenna element $b$ and user $k$.
Finally, we calculate the marginal probability of the symbols of each user by multiplying messages from all the antenna elements and their prior, yielding
\begin{align}\label{eq: final_qx}
    q_{x_k}(x_k) \propto \prod_{b=1}^M m_{f_{y_b}\longrightarrow x_k}(x_k)\times f_{x_k}(x_k).
\end{align}
From the resulting discrete distribution, we can compute the symbols mean and variance as $\mu_{x_k}=\sum_{x_k\in\mathcal{A}}x_k q(x_k)$ and $\mu_{x_k}$ and $\sigma_{x_k}^2 = \sum_{x_k\in\mathcal{A}}|x_k|^2 q(x_k)-|\mu_{x_k}|^2$.
\subsubsection{Damping factor}
In order to improve the convergence of our VMP-based scheme, we use a damping factor $\delta_{vmp}$ to smooth the updates of $q^{(t)}_{x_k}(x_k)$ at $t$th iteration as \cite{gao2017massive}
\begin{align}\label{eq: damping}
    q^{(t)}_{x_k}(x_k) \Leftarrow
    \delta_{vmp}q^{(t)}_{x_k}(x_k)+
    (1-\delta_{vmp})q^{(t-1)}_{x_k}(x_k)
\end{align}
where the symbol "$\Leftarrow$" denotes the assignment and $\delta_{vmp}\in [0,1]$ performs a weighted average over the messages in the current and previous iterations.

\subsection{MRC initialization}
One of the important factors in the convergence rate of the message passing-based algorithms is the initialization. Here, we propose an initial MRC processing over the received signal and feed the soft information to the VMP algorithm for further processing.

In order to apply MRC to the received signal, we need to use $\mathbf{F}_{\text{MRC}}=\frac{\mathbf{h}_k^H}{||\mathbf{h}_k||^2}$ filter on \eqref{eq:general_model} that yields to 
\begin{align}
    x_k^{\text{MRC}}=\frac{\mathbf{h}_k^H}{||\mathbf{h}_k||^2}{y}= x_k + \sum_{k'\neq k}^K\frac{\mathbf{h}_k^H}{||\mathbf{h}_k||^2} \mathbf{h}_{k'} x_{k'}+\frac{\mathbf{h}_k^H}{||\mathbf{h}_k||^2}\mathbf{n}
\end{align}
Assuming the crowded scenario mode ($K\gg 1$), the second and third terms in $x_k^{\text{MRC}}$ 
can be approximated as complex Gaussian random variable according to the central limit theory. Therefore, we set the initial marginal of $x_k$'s as
\begin{align}\label{eq:MRC_init}
    {q^0(x_k)} \!=\! \mathcal{CN}\bigg(\!x_k; x_k^{\text{MRC}}\!\!\!,\frac{\sum_{k'\neq k}^K
    |\mathbf{h}_k^H \mathbf{h}_{k'}|^2 +||\mathbf{h}_k||^2\sigma_n^2}
    {P_{x_k}||\mathbf{h}_k||^4}\!\bigg)
\end{align}
where $P_{x_k}=\mathbb{E}\{x_kx_k^H\}$ is user signal power.

\subsection{Algorithm}
Our proposed receiver is summarized in Algorithm~\ref{alg}, where VMP solver and the XL-MIMO system parameters are given as inputs and detected symbols are the outputs. After generating the true non-stationary MIMO channel and initial symbol probabilities by MRC, several loops in the VMP begin to exchange messages and update the variables. The first loop calculates the different parameters for the desired messages and updates them until a predefined number of iterations $\mathcal{I}$. Finally, the second loop chooses the most probable symbol for each of the users.

\begin{algorithm}[t]
	\SetAlgoLined
	\KwResult{Symbol detection for all  active users}
	\emph{Initialize:} 
	 $M$,  $K$,  parameters in \ref{sec:system model}, $\mathcal{A}$, VMP iterations $\mathcal{I}$,$\delta_{vmp}$.
	
1. Generate channel matrix $\mathbf{H}$ using \eqref{eq:ch_model}.

2. Calculate the initial MRC probabilities $q^0(x_k)$ using \eqref{eq:MRC_init}.

\For{$i = 1$ to $\mathcal{I}$ }{

3. Extract $\mu_x$ and $\sigma^2_x$ values from  $q_{x_k}^{(i-1)}(x_k)$. 

4. Calculate the mean value of the precision parameter $\hat{\lambda}_b$ using \eqref{eq:lambda_bar} for all the antenna elements $b=\{1,\cdots, M\}$.

5. Calculate symbol probabilities $q^{(i)}_{x_k}(x_k)$ using \eqref{eq: final_qx} for all the users $k=\{1,\cdots, K\}$.

6. Update the symbol probabilities applying the damping factor in \eqref{eq: damping}.
		
	}
\For{$k = 1$ to $K$ }{
7. $\bar{x}_k=arg\max_i q^{(\mathcal{I})}_{x_k}(x_k=a_i|a_i\in \mathcal{A}) $

}

	\caption{\small VMP with MRC initialization.}
	\label{alg}
\end{algorithm}

\section{Simulation Results}
\begin{figure}
	\centering
	\includegraphics[width=1\linewidth]{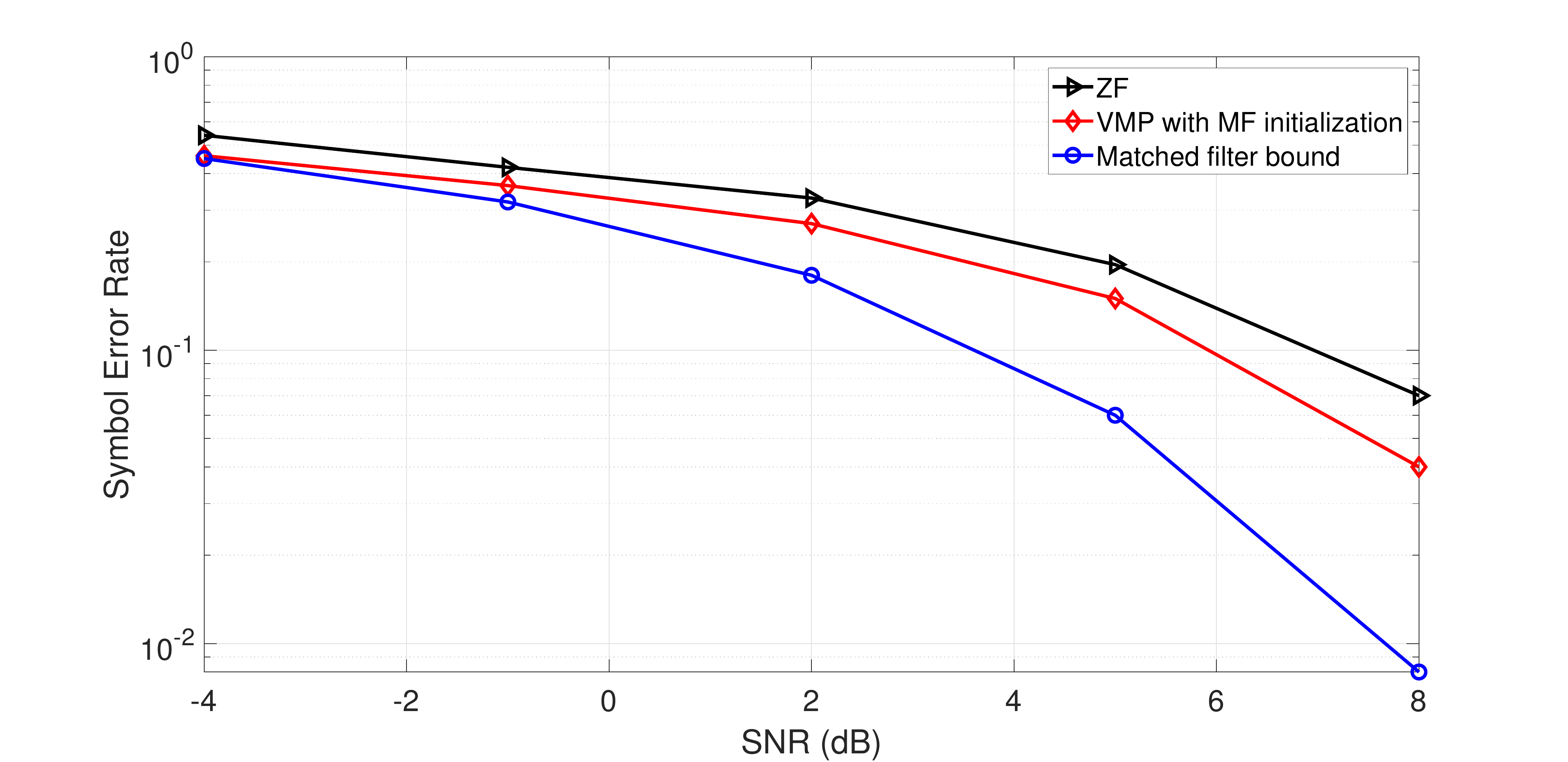}
	\caption{ \small SER comparison of the different detection methods in a heavily loaded XL-MIMO system with $\frac{M}{K}=2$
	}
	\label{fig:sim1}\vspace{0.4cm}
\end{figure}

In this section we evaluate the performance of the proposed algorithm and compare it with the other benchmark methods. 
{We choose an ideal bound where perfect interference removal is done for each target user and then MRC is used for single-user detection in the interference-free channel.}
 We call this bound ``matched filter bound"\cite{ClaudeTurbo}.
All of the simulation parameters are shown in Table \ref{tab}. 
According to the numerical analyses VMP in our model converges at most at $\mathcal{I}=3$. 
\begin{table}
\centering
\caption{Simulations parameters in detail.}
\label{tab}
\begin{tabular}{|c|c||c|c|}
\hline
\bf{Variable} & \bf{Value} & \bf{Variable}& \bf{Value} \\\hline	

$M$    &    $512$  &   $K$   &  $256 $  \\\hline

$\mathcal{I}$    &    $3$  &   $|\mathcal{A}|$   &  $4 $  \\\hline

$\mathbf{P}$    &    $\mathbf{I}$  &   $r$   &  Uniform$(5,10) $  \\\hline

$L$    &     $29.51 m$  &   $\nu$   &  $3 $  \\\hline

$\lambda$    &    $2.6$GHz  &   Antenna spacing   &  $\lambda/2 $  \\\hline

$(\mu_l,\sigma_l)$ & $(2.25,0.1)$  &  $\zeta$   &  $M/4 $  \\\hline

$\Omega$    &    $4$  &   $\delta_{vmp}$   &  $0.45 $  \\\hline

\end{tabular}

\end{table}
Fig.~\ref{fig:sim1} shows the SER comparison of the proposed method, ZF and the ideal single user bound. As it can be seen, the VMP based algorithm outperforms the ZF detector while keeping an acceptable gap with the ideal bound. As mentioned before, due to lack of the favorable propagation, ZF fails to work near-optimally.

\begin{figure}
	\centering
 	\includegraphics[width=1\linewidth]{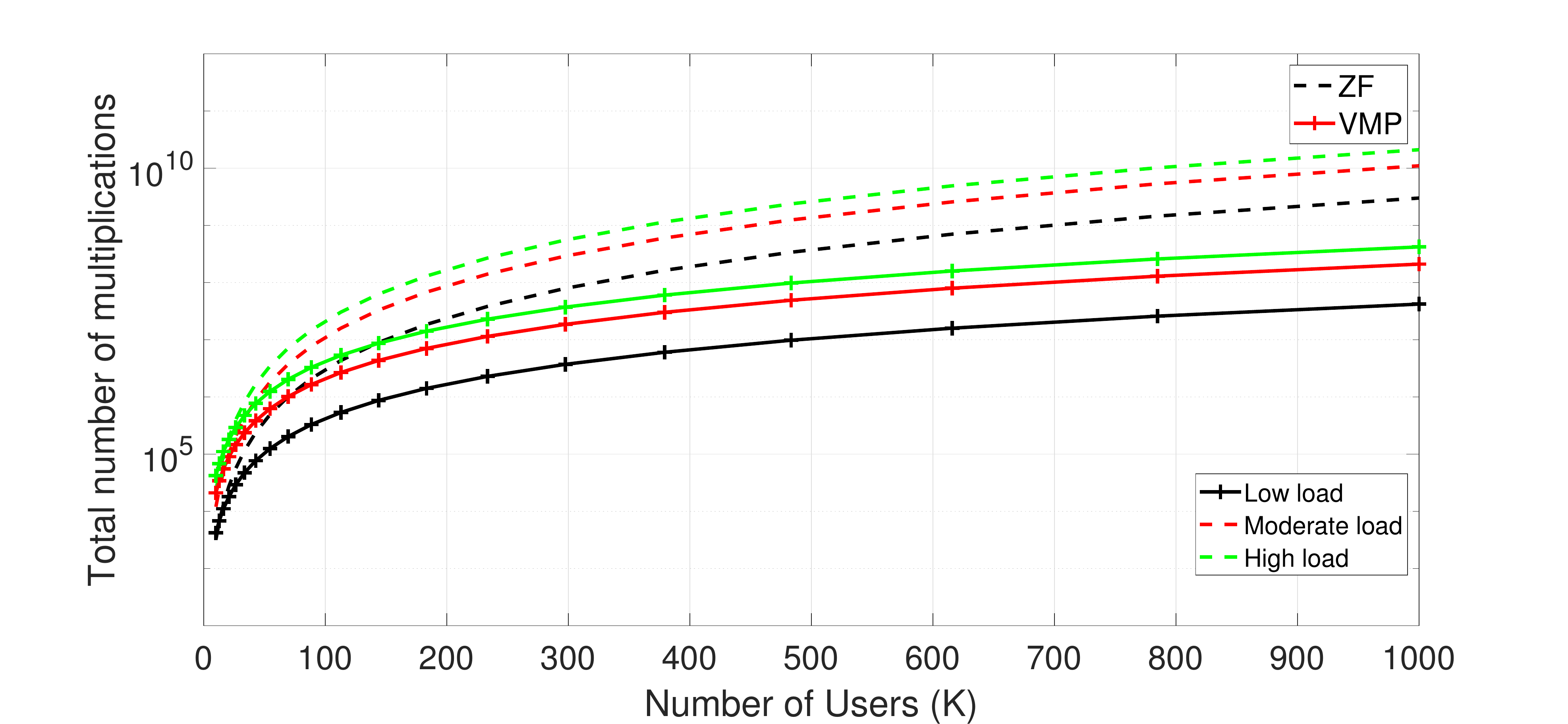}
	\caption{ \small Complexity comparison of the different detection methods in three system load modes.
	}
	\label{fig:sim2}\vspace{0.4cm}
\end{figure}
\subsection{Complexity Analyses}
Here, we derive the computational complexity of the proposed method and the benchmark method zero-forcing (ZF). The complexity of ZF is \cite{bjornson2015optimal}
\begin{align}
    C_{\mbox{ZF}}=\frac{K^3}{3}+MK^2+MK
\end{align}
while the complexity of the VMP-based method is
\begin{align}
    C_{\mbox{VMP}}=\mathcal{I}(\underbrace{M(3+2K)}_{(I)}+\underbrace{MK|\mathcal{A}|}_{(II)})+\underbrace{3MK}_{(III)}
\end{align}

where, in $(I)$ we have 3 multiplications for updating $\mu_b$, $\sigma^2_b$ and $\hat{\lambda}_b$ for each of the antennas plus 2 multiplications per user for deriving $\mu_x$ and $\sigma^2_x$. Then $(II)$ stands for executing \eqref{eq: final_qx} and finally $(III)$ is for the MRC initialization part in \eqref{eq:MRC_init}.

Fig.~\ref{fig:sim2} compares the complexity of these two methods in three different scenarios of high, moderate and low load regimes with $M/K$ equal to $2$, $10$ and $20$, respectively. The  total number of the multiplications for VMP is always smaller than the ZF's and the gap grows as we approach to the crowded scenarios with much more users in the system.

\section{Conclusions}

In this work, we have studied the design of multi-user detection schemes for XL-MIMO systems. We have shown that VMP can be used to design a message-passing receiver with complexity that scales linearly with the number of users and antenna elements, thus making it suited for XL-MIMO systems with high system load. In addition, the detection performance surpasses that of a classical ZF detector, in spite of requiring fewer computations. Future research will address the inclusion of channel estimation together with data detection in the VMP receiver and the exploration of distributed implementations.


 \bibliographystyle{IEEEtran}

\end{document}